# Electric and Magnetic Field Nano-Sensing Using a New, Atomic-like Qubit in a Carbon Nanotube

I. Khivrich[1] and S. Ilani[1*]

[1]Department of Condensed Matter Physics, Weizmann Institute of Science, Rehovot 76100, Israel.

[*] Correspondence to: shahal.ilani@weizmann.ac.il

**Quantum sensing techniques have been successful in pushing the sensitivity limits in numerous fields, and hold great promise for scanning probes that study nano-scale devices and novel materials. However, forming a nano-scale qubit that is simple and robust enough to be placed on a scanning tip, and sensitive enough to detect various physical observables, is still a great challenge. Here we demonstrate a conceptually new qubit implementation in a carbon nanotube that achieves these requirements. In contrast to the prevailing semiconducting qubits that use electronic states in double quantum dots, our qubit utilizes the natural electronic wavefunctions in a single quantum dot. Using an ultra-clean nanotube we construct a qubit from two wavefunctions with significantly different magnetic moments and spatial charge distributions, making it sensitive to both magnetic and electric fields. We use an array of gates to directly image these wavefunctions and demonstrate their localized moments. Owing to their different spatial structure, these wavefunctions also show radically different transport properties, giving us a simple transport-based qubit readout mechanism. Due to its narrow coherence-limited transition, the qubit demonstrates significantly better electric field detection sensitivity than a single electron transistor. Moreover, with the same qubit we demonstrate simultaneous probing of magnetic fields with DC sensitivity comparable to that of NV centers. Our technique has minimal requirements for device complexity, which can be implemented using a number of straightforward fabrication methods. These features make this atomic-like qubit a powerful new tool that enables a variety of new nanoscale imaging experiments.**

Ultrasensitive nanoscale detectors of electric and magnetic fields take an increasingly central role in advancing the research of novel devices and materials. When used as scanning probes, these detectors provide a unique insight to electronic and spin systems on the nanoscale. To date, a large variety of scanning probe sensors have been developed, optimized to measure specific physical quantities: Magnetic fields are primarily imaged via scanning SQUIDs[1], Hall probes[2] and NV centers[3], while electric fields are primarily probed with Kelvin probes[4], scanning tunneling potentiometry[5], and scanning single electron transistors (SET)[6]. From all the above techniques, only NV centers utilize a quantum two-level system (qubit) that takes full advantage of the power of quantum manipulations. This gives NV center-based probes unprecedented sensitivity to local magnetic fields, and additionally a modest sensitivity to electric fields[7]. NV scanning probes have also few limitations: optical readout introduces a significant challenge at cryogenic temperatures, and using the sensor at high magnetic fields requires impractical RF frequencies. A different type of a scanning qubit that can sense electric fields ultra-sensitively on the nanoscale and simultaneously probe magnetic fields with modest sensitivity, will have complementary capabilities and is thus highly desirable.

Candidate solid state qubits for nanosensing applications generically divide into two groups – atomic and engineered. Atomic qubits (such as NV centers[8] or P dopants in Si[9,10]) utilize natural atomic wavefunctions as their basis, and hence are small and often have long coherence times. Engineered qubits, on the other hand (e.g. transmon[11] or semiconducting double quantum dot qubits[12,13]) provide finer control over the energy spectrum and the dipole coupling to the physics of interest, however, they are larger ($\mu m$s to $mm$s), require complex planar circuit designs, and often rely on external detectors for their readout, thus making them less suitable for nanoscale sensing applications. A qubit that can combine the simplicity of atomic qubits with the tunability and control of engineered qubits could therefore lead to a new and potentially powerful scanning nanosensor.

A conceptually simple qubit that may combine the above advantages can be based on the natural electronic wavefunctions of a single quantum dot. Similar to atomic orbitals, such wavefunctions have distinct spatial structure. This structure, however, occurs on much larger spatial scales and can therefore provide larger and more tunable electric moments. Carbon nanotubes present an excellent setting for realizing this concept; in their recent generations they are electronically pristine, allowing the creation of quantum dots with exceptional level of control over their wavefunctions and energy spectrum[14,15]. So far, engineered double quantum dot qubits have been demonstrated successfully in carbon nanotubes[16–19], but the lithographic complexity of these devices and their frequent reliance on extra readout such as on-chip resonators, may be prohibitive for using them in scan probes. Moreover, most of these qubits were intentionally designed to be insensitive to external fields and thus are poor sensors. At the same time, single quantum dot devices in carbon nanotubes have been successfully used as ultra-sensitive scanning SETs, allowing to image oxide interfaces[20], Wigner crystals[21], as well as the mapping of ballistic[22] and hydrodynamic[23] electron flows, demonstrating the compatibility of these devices with scanning probe applications.

In this work, we realize a new type of a qubit in a carbon nanotube, and demonstrate its application as a highly-sensitive electric and magnetic fields nano-sensor. Our qubit has a simple built-in transport-based readout, a highly local response to electric fields, which we image directly using capacitive techniques, and simplicity that allows placing it at the edge of a scanning probe cantilever. We determine its decay and dephasing times using time-domain and Landau-Zener-Stuckelberg[22] interferometry experiments and show that its coherence-limited transition leads to significantly improved electric potential sensitivity as compared to the thermally-broadened Coulomb blockade peak of an SET. Furthermore, we demonstrate that the same qubit can simultaneously detect magnetic fields parallel to the nanotube axis. Although the short coherence time did not allow us to implement dynamic decoupling protocols and compete with the AC sensitivity of NV centers, we achieve DC magnetic field sensitivity that is on par with that of NV center[24] and Hall bar-based[25] scanning probes.

The basis of our qubit is given by two electronic wavefunctions in a single quantum dot, formed in a suspended carbon nanotube. Within the single-particle picture, a parabolic confinement potential along the nanotube leads to a ladder of harmonic oscillator levels, whose wavefunctions' extent along the nanotube axis increases with increasing level number (Fig 1a, gray illustrations). Each level is 4-fold degenerate, due to the spin (↑,↓) and valley ($K, K'$) degrees of freedom. In a gapped nanotube, $K$ and $K'$ electrons rotate in opposite directions around the nanotube circumference, leading to opposite orbital magnetic momenta. Applying a magnetic field parallel to the tube axis, $B_{||}$, breaks the spectrum into four independent ladders with slopes given by the orbital and spin magnetic moments, $\frac{\partial E}{\partial B_{||}} = \pm\mu_{spin} \pm \mu_{orb}$, with $\mu_{orb} \gg \mu_{spin}$ (red and blue lines in Fig.1a correspond to $K$ and $K'$ states). Spin-orbit coupling splits[26] the 4-fold degeneracy at $B_{||} = 0$, and Coulomb repulsion changes the simple non-interacting wavefunctions into Wigner crystals with finer real-space structures[21], yet, since the valley remains a good quantum number, the simple picture in which $B_{||}$ leads to crossing between levels with different magnetic moments and different spatial structures remains valid for the discussion below.

To make an atomic-like qubit that is sensitive to local electric and magnetic fields, we choose a crossing between a high-lying $K$ state and a low-lying $K'$ state with opposite spin directions, which we will denote as $K_n$ and $K'_m$ ($n \gg m$). The charge density of the $K_n$ state is spatially extended (Fig 1b. left, red) whereas that of the $K'_m$ state is spatially localized (Fig 1b. left, blue), endowing the qubit transition a localized electric moment. Contrary to a standard charge qubit in double quantum dots, whose charge is localized on the left or right dots, separated by lithographic dimensions, in our case the electrical moment results from the difference in charge distribution of different wavefunctions within a single quantum dot. Due to symmetry of the charge distributions around the center of the dot, the dipole moment of the qubit is approximately zero, potentially reducing its sensitivity to homogenous electric fields and far-field noise, however, this qubit has quadrupole or higher moments that yield strong sensitivity to local fields, which is beneficial for high resolution imaging. The $K_n$ and $K'_m$ states have also opposite orbital momenta (Fig. 1b, center), endowing the qubit transition also a large magnetic moment ($\sim 20\mu_B$, $\mu_B$ is the Bohr magneton). We choose opposite spins for the basis states to minimize their overlap, which leads to long decay times.

The schematic charge stability diagram for a hole-doped nanotube single quantum dot is plotted in Fig. 1c as a function of gate voltage, $V_G$, and $B_{||}$. The diagram is obtained by adding the charging energy $U$ to single-particle energies in Fig. 1a, inset. Transport occurs along Coulomb blockade (CB) charging lines, across which a hole is added to the system, zig-zaging between $K$ (red) and $K'$ (blue) character as a function of $B_{\parallel}$. The relevant triple point for our experiment separates the state $|N\rangle$, having $N$ holes, from the two qubit states, $|B\rangle = |N\rangle + K_n$, and $|D\rangle = |N\rangle + K'_m$, both having $N+1$ holes. Within the Coulomb valley there should be a boundary line separating the $|B\rangle$ and $|D\rangle$ ground states (dashed green) along which the ground state changes its last occupied wavefunction while maintaining the total charge in the dot fixed. This line will however be invisible in transport, as the system is in Coulomb blockade.

We implement the atomic-like qubit in a device that has an single carbon nanotube, nano-assembled[15] at the edge of a cantilever (Fig. 1d). The nanotube is suspended a distance of $1.2\mu m$ between two contacts (S,D) over an array of seven individually controlled gates (Fig. 1d, inset). Note that this geometry is identical to our scanning nanotube-based SET cantilever geometry, which we previously used to image 1D[19] and 2D[20,23,27] systems, making the technique developed here directly applicable for scanning probe applications. The device is cooled in a dry dilution refrigerator, with an electron temperature of $T_{el} \sim 60mK$ as measured by the width of CB peaks, and with a magnetic field parallel to the nanotube axis. The nanotube conductance, $G$, is measured at zero $V_{sd}$ bias using an $LC$ tank circuit connected to the drain contact[28], and with a small AC excitation on source contact ($\sim 15\mu Vrms$) at the tank circuit resonant frequency (Supp. Info S1).

Fig. 1e shows $G$, measured as a function of a common gate voltage, $V_G$, applied together on all gates, and $B_{\parallel}$. The expected zig-zag behavior of the CB peaks is clearly visible, however, while the $K$ transitions exhibit finite conductance at the CB peak ('bright'), the $K'$ transitions have no observable conductance ('dark'), and are marked in the figure by dashed blue lines. Similar dark/bright behavior of the two valleys at finite $B_{\parallel}$ was observed previously[29,30]. The triple point used for our experiments is shown in the zoom-in measurement (Fig. 1f) with the three relevant ground states, $|N\rangle$, $|B\rangle$ and $|D\rangle$, labeled. To clarify, the 'dark' states described here are not equivalent to the recently reported dark states due to coherent population trapping[31], which occur at finite $V_{sd}$, and do not require $B_{\parallel} > 0$.

At finite $B_{||}$, the $p-n$ junction barriers that confine the holes in the two valleys differ significantly (Fig 1b right). Their height and spatial extent, given by the nanotube bandgap, decreases with $B_{||}$ for the $K$ states ($dE_{gap}^{K}/dB_{||} = -2\mu_{orb}$) and increases with $B_{||}$ for the $K'$ states ($dE_{gap}^{K'}/dB_{||} = 2\mu_{orb}$), leading to markedly different transport for the two states. The difference in transport visibility between the $|N\rangle \leftrightarrow |B\rangle$ and $|N\rangle \leftrightarrow |D\rangle$ transitions thus gives a built-in transport-based readout mechanism for the qubit state, which does not require an external charge detector.

In order to form a charge qubit, sensitive to its electric environment, its basis states ($|B\rangle$ and $|D\rangle$) should differ in their charge distribution along the nanotube. We image these charge distributions directly using the array of gates, as follows; First, we tune the voltage common to all gates, $V_G$, to observe the Coulomb peak at the $|N\rangle \leftrightarrow |B\rangle$ transition (illustrated in gray, Fig 2a bottom). Then, we repeat this scan but with a voltage offset $\Delta V$ added to gate $i$. This will lead to a shift the Coulomb blockade peak by $\delta V_i$ along the $V_G$ axis, proportional to the local charge density just above this gate (Colored curves in Fig.2a bottom, Supp. Info S2). By measuring the individual shifts with respect to all gates, $\delta V_i$, $i = 1..7$, we thus image the charge density added on the transition at seven spatial points, which is essentially the discrete version of the scanning imaging of Wigner crystals that we performed previously[21]. Although the $|N\rangle \leftrightarrow |D\rangle$ transition is dark in transport, we still know its position accurately by connecting the corners of the bright transitions (dashed blue, Fig 1f). Thus, using the same method we can image also the spatial charge density within the $|D\rangle$ state.

Figure 2b zooms in on the triple point around $|N\rangle$, $|B\rangle$ and $|D\rangle$ (white square in Fig. 1f). Upon addition of $\Delta V = 0.5mV$ to gate 4 the bright and dark transition shifts by independent amounts, $\delta V_4^B$ and $\delta V_4^D$ (gray arrows). Similar measurements with all gates yields the shifts $\delta V_i^B$ and $\delta V_i^D$, which when plotted as a function of gate position (Fig. 2c and 2d) trace the spatial distribution of the charge added at the $|N\rangle \leftrightarrow |B\rangle$ and $|N\rangle \leftrightarrow |D\rangle$ transitions, $\rho_{NB}(x)$ and $\rho_{ND}(x)$, where $x$ is the spatial coordinate along the nanotube. Visibly, while $\rho_{NB}(x)$ is homogenously spread over all gates, $\rho_{ND}(x)$ is localized at the dot's center.

To study the dynamics of a $|D\rangle, |B\rangle$ qubit we turn to time domain experiments that use gate voltage on the central three gates, $V_G$, as a fast control axis, with the following sequence: First, the dot is initialized in the $|B\rangle$ state, on the $|N\rangle \leftrightarrow |B\rangle$ Coulomb peak ($V_G = V_{CB}$, black star, Fig. 3a). Then, a fast ramp to $V_G = V_{probe}$ is applied, after which the system is left to evolve for time $\tau_{probe}$. Finally, the voltage is swept back to the initial CB point for readout, dwelling for time $\tau_{read,init}$. If after the probing stage the system ended up in the ground state $|B\rangle$, the dot will freely conduct in the readout stage. However, if the system switched to the excited state $|D\rangle$, it will remain in the dark state during readout, blocking the conductance. The characteristic blocking time is given by the fastest of two possible decay routes, $|D\rangle \to |N\rangle$ or $|D\rangle \to |B\rangle$, both of which initialize the system to its ground state. The above sequence is repeated periodically, and we measure the conductance averaged over this sequence, which contains two terms: $\langle G \rangle = \left(G(V_{probe})\tau_{probe} + G(V_{CB})\langle P_B\rangle \tau_{read,init}\right)/(\tau_{probe} + \tau_{read,init})$. The first term reflects the conductance measured during the probing stage, and is non-zero only for $V_{probe}$ near the Coulomb peak, where the dot has a finite conductance. The second term reflects the conductance measured in the readout stage, and is directly proportional to the mean bright state probability, $\langle P_b \rangle$ during this stage.

Figure 3b shows $\langle G \rangle$ measured as a function of $V_{probe}$ within the above sequence (red), using $\tau_{probe} = 0.8 \mu s$ and $\tau_{read,init} = 5\mu s$, as well as the measured quasi-DC conductance, $G$, (blue). The CB peak in $G$ appears also in $\langle G \rangle$, as expected from the first term the equation above. Interestingly, however, inside the Coulomb valley $\langle G \rangle$ shows a sharp dip at $V_{probe} = V_{BD} \approx -343 mV$, not present in $G$. This dip is much narrower ($\sim 40 \mu V$) than the thermally-limited CB peak ($\sim 200 \mu V$). From $\langle P_b \rangle$ extracted from $\langle G \rangle$ and $G$ using the equation above (Fig. 3c) we see that for most values of $V_{probe}$ the state remains bright ($P_B = 1$), but at the dip the dark state becomes significantly occupied ($\langle P_B \rangle \approx 0.7$). Repeating the above measurement at various values of $B_{||}$ (Fig. 3d) shows that this dip traces a straight line terminating at the $|N\rangle$, $|B\rangle, |D\rangle$ 'triple point', as expected from the $|B\rangle \leftrightarrow |D\rangle$ degeneracy line (dashed green in Fig. 1c.). Its finite slope suggests that this transition is sensitive to both local magnetic and electric fields, where the latter attests to the different charge distribution within the $|B\rangle$ and $|D\rangle$ states.

Similarly to Fig.2, we can image directly the charge density distribution change at the $|B\rangle \leftrightarrow |D\rangle$ transition, by measuring the response of the transition line position, $V_{BD}$, to small gate perturbations. The measured density distribution, $\rho_{BD}(x)$ (Fig. 3e, left) compares well to difference between the bright (Fig. 2c) and dark (Fig. 2d) state densities, $\rho_{BD}(x) \approx \rho_{NB}(x) - \rho_{ND}(x)$ (Fig. 3e, right), further establishing the narrow transition line as the boundary between the $|B\rangle$ and $|D\rangle$ ground states. From the measured $\rho_{BD}(x)$ we see that the qubit charge redistribution is also narrow in space. The observed width (~200nm, Fig. 3a) is limited by the resolution of the imaging method due to the size and distance to the gates. A more quantitative analysis that deconvolves the known shape of the potential distribution produced by the gates concludes that the actual width is ~100nm (Supp. Info. S3). This width sets the spatial resolution of the qubit sensor.

To measure the transition rate ($T_1$ time) and its dependence on the detuning from the degeneracy point, $V_{BD}$ (Fig.4a), we repeat the measurements above but with different dwell times, $\tau_{probe}$. In Fig. 4b we plot $\langle P_B \rangle$ as a function of the voltage offset, $\Delta V_G = V_{probe} - V_{BD}$, for various $\tau_{probe}$ values, and in Fig. 4c we plot it as a function of $\tau_{probe}$ for different values of $\Delta V_G$. Away from the dip, the decay time is too long to reliably be extracted from this figure, whereas at the dip it becomes significantly shorter, $T_1 \sim 1\mu s$, indicating a fast transition from $|B\rangle$ to $|D\rangle$.

To observe quantum coherence of the $|D\rangle, |B\rangle$ qubit and estimate its $T_2^*$, we use Landau-Zener-Stuckelberg (LZS) interference[22,32]. In this case, instead of waiting at $V_{probe}$ for $\tau_{probe}$, the detuning is steered as $V_G(t) = V_{BD} + \Delta V_G + A_{LZS} \sin(2\pi f_{LZS} t)$ (Fig. 4d) and we probe $\langle P_B \rangle$ after time $\tau_{probe}$ by moving $V_G$ to the CB peak and measuring $\langle G \rangle$ as before. The $\langle G \rangle$ measured as a function of $\Delta V_G$ and $A_{LZS}$ at a frequency of $f_{LZS} = 0.7 GHz$ (Fig.4e) shows the characteristic LZS interference pattern. The peak width, $\delta \omega \sim 2\pi \times 180 MHz$, indicates that the qubit maintain coherence over several oscillations, having a $T_2^*$ time of ~0.9ns.

The observations can be quantitatively explained by a simple model, describing the evolution of the system in the $|D\rangle, |B\rangle$ manidfold; The unitary evolution is described by the Hamiltonian $H = \epsilon(t)\sigma_z + \Delta \sigma_x$, where the $|D\rangle$ and $|B\rangle$ are the eigenvectors of $\sigma_z$.

The dominant decoherence mechanism with rate $\gamma_2$ results from coupling to charge noise acting along the energy detuning axis, $\epsilon(t)$, coupling only to $\sigma_z$. In the far-detuned regime ($\epsilon \gg \Delta, \gamma_2$), the noise changes the phase difference between the basis states ($T_2$ processes), however, close to zero detuning it translates to incoherent transition rate between the basis states ($T_1$ process) (Supp. Info. S4). Consequently, the transition rate between the states depends sharply on the detuning (Fig. 4a, bottom). Using this model we quantitatively fit the results in Fig. 4b,c (lines) and obtain the qubit's splitting, $\Delta = 2\pi \times 2MHz$, and its decoherence rates $\gamma_1 = 2\pi \times 1.5kHz$, $\gamma_2 = 2\pi \times 185MHz$ (details in Supp. Info S5), where $\gamma_1$ results from the noise coupled through $\sigma_x, \sigma_y$. With the same parameters we also reproduce (Fig. 4f), quantitatively well, the LZS measurements in Fig 4e (Simulation details in Supp. Info S10).

The strong dependence of the qubit transition on electric potential and $B_\parallel$ implies that it can serve as an excellent nano-scale probe of these quantities. Since SET is the most sensitive scanning electrometer to date, we benchmark the qubit sensitivity against measurements in SET modality. DC electric potential sensitivity is measured by slowly ramping the central gate voltage ($V_4$) up and down, and monitoring conductance changes in the two modalities: On the qubit $|B\rangle \leftrightarrow |D\rangle$ transition line using fast gating, and on the SET $|N\rangle \leftrightarrow |B\rangle$ CB line within the same triple point. Parameters are optimized separately for each modality, and the results are converted to a common potential scale (Fig. 5a). Visibly, the qubit provides a significantly improved sensitivity, primarily due to its sharper transition line as compared to the temperature-limited Coulomb blockade peak. The sensitivity to detuning that we obtain in the qubit measurements is $\sim 60 neV/\sqrt{Hz}$, which translates to a potential sensitivity of $\sim 600 nV/\sqrt{Hz}$ (Supp. Info S7), significantly improving over the performance of the device as an SET, and surpassing the sensitivity of our best SETs to date[27]. DC $B_\parallel$ sensitivity is measured in a similar fashion (Fig. 5b). Here, the advantage of qubit detection as compared to an SET becomes even more evident, reaching a sensitivity of $\sim 39 \mu T/\sqrt{Hz}$ (Supp. Info S8), comparable to the DC magnetic field sensitivity of NV centers[24]. What limits the sensitivity in the current experiment is the large contact resistance of our device ($R_c \sim 2M\Omega$) and unavoidable magnetic field fluctuations inherent to a magnet power supply. Theoretical estimates predict that the performance can be improved by more than an order of magnitude by improving the contact resistance and using a persistent mode magnet (Supp. Info. S6).

Our sensor requires finite $B_\parallel$ for its operation, however, it can operate in a wide range of magnetic fields (demonstrated at $3-8T$, see Supp. Info S9). This provides complementary capability to that of scanning SQUIDs and NV centers, which generally work only at lower fields, although achieving better magnetic field sensitivities. The spatial resolution demonstrated here ($\sim 100nm$) was limited by the rather long ($1.2\mu m$) device used in this study to enable the gate imaging in Fig. 2. In principle, it should be straightforward to implement the same qubit in a much shorter and simpler, single-gated suspended device and the resolution will scale in proportion, to the tens of $nm$ range. The geometry of the current device is equivalent to the standard scanning SET cantilevers, and is thus fully compatible with scanning.

In summary, we have demonstrated a new type of qubit in carbon nanotubes that combines the advantages of atomic and engineered qubits. This qubit is conceptually simple, requires only conventional fabrication, has a small form factor allowing placing it on a scanning probe tip, has a simple built-in readout mechanism, and enables sensitive measurements of electric and magnetic fields. These features make it an enabling tool for a variety of experiments. For example, since an atomic-like qubit can be much smaller than lithographic dimensions, which constrain double-dot qubits, it can couple to high vibrational modes of suspended carbon nanotubes, which are at their quantum ground state at dilution temperatures, thus enabling quantum nano-mechanical experiments in this system. As a scanning detector that measures simultaneously electric and magnetic fields, it will be instrumental in exploring phenomena that have both charge and magnetic (/electric current) signatures. Few examples include imaging current whirlpools in hydrodynamic electron flow[33], electrical charges of topological magnetic structures[34], and resolving the microscopic interplays in multiferroics[35]. More broadly, the addition of quantum sensing and time domain capabilities into scanning electrical field measurements opens the door for sensitivity improvements and for imaging the dynamics in quantum systems, that were so far beyond reach.

**Acknowledgements:** We thank A. Finkler, B. Kalisky, F. Kuemmeth and E. Zeldov for helpful suggestions. We further acknowledge support from the Minerva grant no.

712290, the Helmsley Charitable Trust grant, and the ERC-Cog (See-1D-Qmatter, no. 647413).

**Data availability:** The data that support the plots and other analysis in this work are available from the corresponding author upon request.

**Contributions:** I.K. and S.I. conceived the experiment. I.K. built the experimental system, performed the experiments, analyzed the data and preformed the simulations. I.K. and S.I. wrote the manuscript.

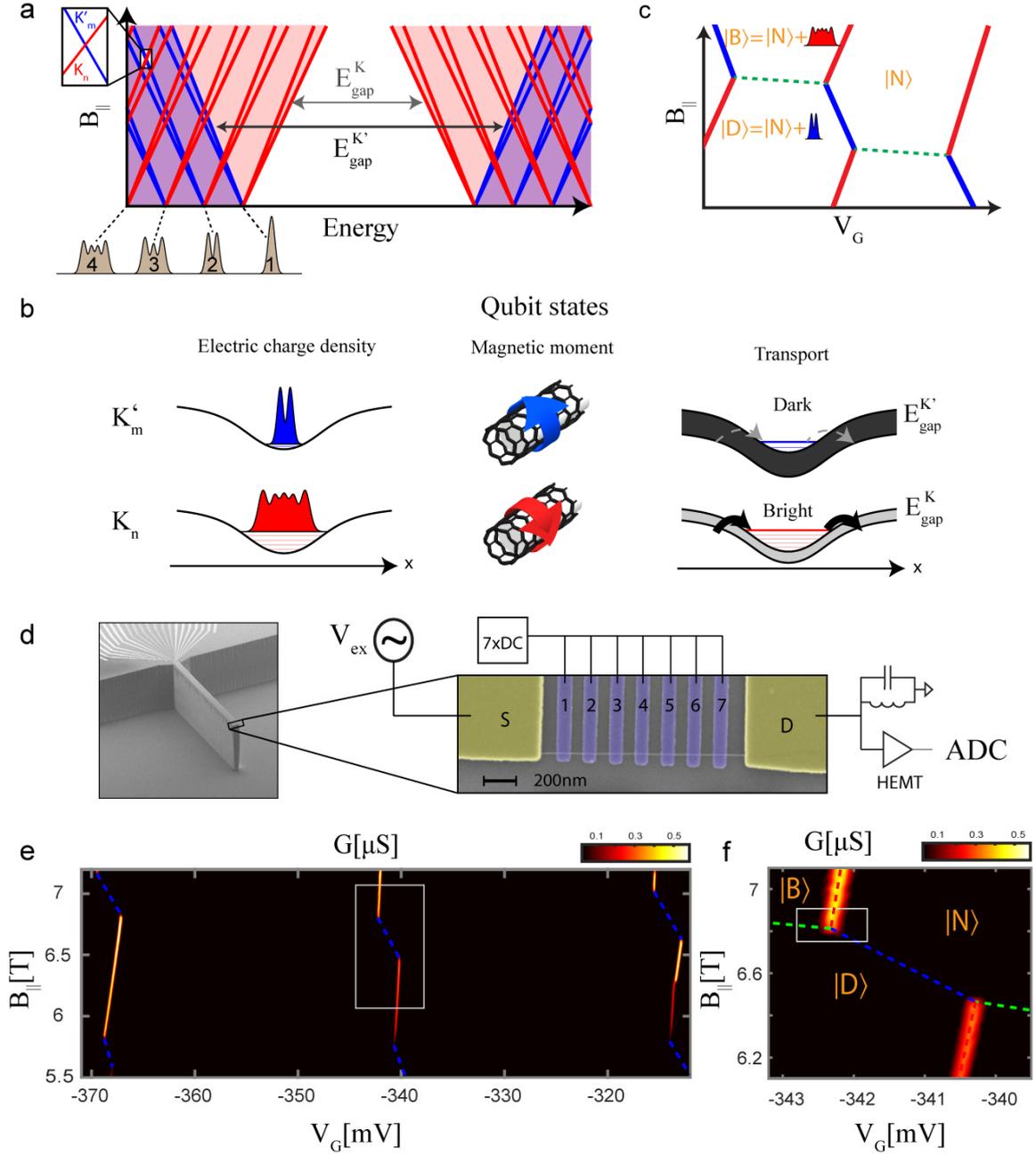

**Figure 1: Natural wavefunctions qubit in a carbon nanotube. a.** Single particle energy spectrum of a single quantum dot in a carbon nanotube, as a function of magnetic field parallel to the tube axis, $B_{\|}$. Each spatial wavefunction (gray illustrations, bottom) has at $B_{\|} = 0$ a 4-fold spin ($\uparrow, \downarrow$) and valley ($K - red, K' - blue$) degeneracy, which is lifted at finite $B_{\|}$ with slopes given by the spin and orbital magnetic moments $\pm\mu_{spin} \pm \mu_{valley}$. $B_{\|}$ also modifies the bandgap of the two valleys: $E_{gap}^{K}$ decreases and $E_{gap}^{K'}$ increases with $B_{\|}$ (gray arrows). The intersection of two levels at finite $B_{\|}$ (inset) is used as the basis of our qubit. **b.** The two intersecting levels (referred to as $K_n$, $K'_m$) differ in three quantities: Since the $K_n$ state is a much higher bound state in the confinement potential than the $K'_m$ state, it is spread more along the nanotube (left). The $K_n$ and $K'_m$ states originate from opposite valleys with opposite magnetic moments due to opposite directions of electron motion around the nanotube circumference (center). The barriers of the dot are formed by the nanotube bandgap, which at finite $B_{\|}$ is different for the two valleys

(see panel a), making the tunneling from the leads intothe $K_n$ state much faster than to the $K'_m$ state ('bright'/'dark', right panel). **c.** Charge stability diagram, obtained by adding the charging energy to the energy spectrum in panel a, inset. Coulomb blockade peaks zig-zag between charging of the two valleys (red/blue). The triple point used in the experiment is between the $|N\rangle$ state, having N holes, and the $|D\rangle$ and $|B\rangle$ states, which are obtained by adding either a $K_n$ hole (red) or a $K'_m$ hole (blue) **d.** Scanning electron microscope image of the device: At the edge of an etched cantilever, a nanotube is positioned on two contacts (S,D) and suspended over a piano of seven gates. The conductance through the device is measured with an AC excitation on S ($V_{ex}$ at $\sim 1.5 MHz$) and a cryogenic LC tank circuit connected to D followed by a cold HEMT amplifier. **e.** Measured conductance, $G$, as a function of gate voltage common to all seven gates, $V_G$, and $B_{||}$, exhibiting a zig-zag of bright and dark charging lines (the latter marked by dashed blue). **f.** Zoom-in on one triple point, with the schematic lines separating the $|N\rangle$, $|D\rangle$ and $|B\rangle$ states are marked as in panel c.

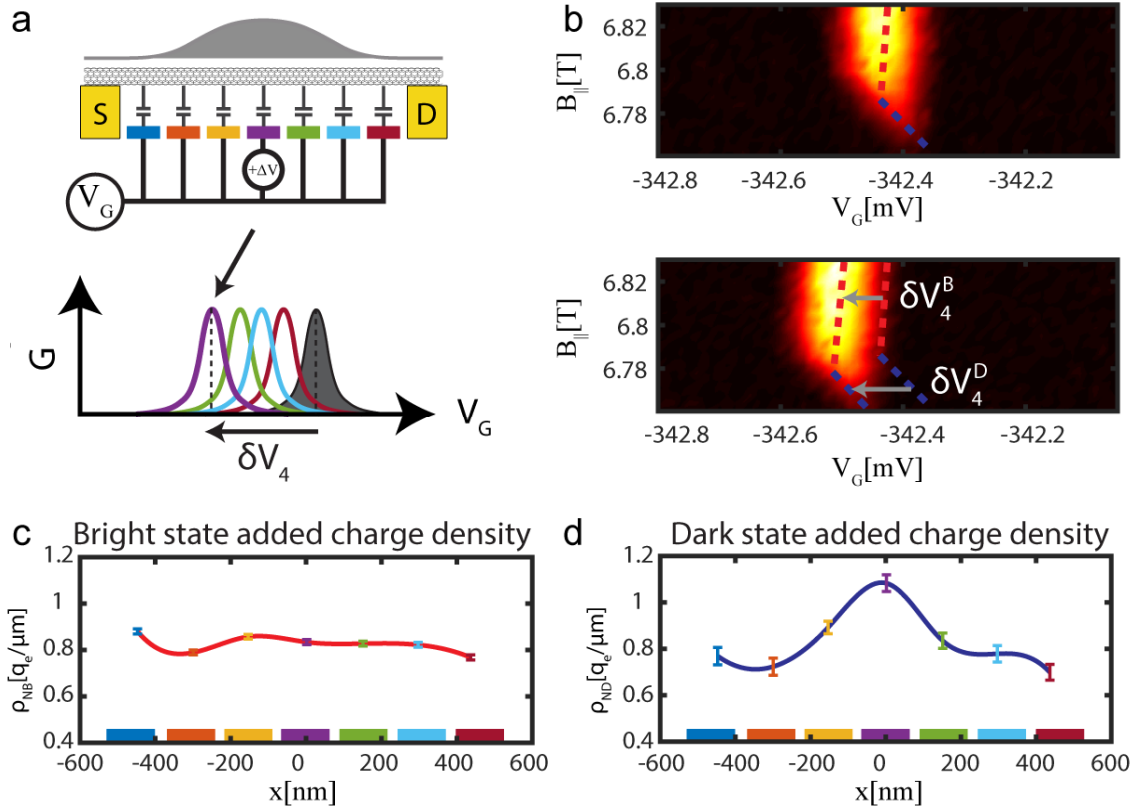

**Figure 2: Imaging the bright and dark states' spatial charge distributions. a**. The charge distribution of a state is imaged by measuring its independent capacitances to each of the individual gates (colored). This is achived by monitoring how the state's Coulomb blockade charging peak (gray, bottom) shifts in common gate voltage, $V_G$, when a small voltage perturbation, $\Delta V$, is added to each of the gates (colored peaks). The individual shifts are directly proportional to the local charge densities above the corresponding gates. **b.** Top: Zoom-in on the triple point in the white box in Fig. 1f. Bottom: Same triple point when a perturbation $\Delta V = 0.5 mV$ was added to gate 4. Bright and dark states charging lines (dashed red and blue) have independent shifts along $V_G$, labeled $\delta V_4^B$ and $\delta V_4^D$. **c.** Measured spatial distribution of the bright state added charge, $\rho_{NB}(x)$, where $x$ is the spatial coordinate along the nanotube. Colored errorbars correspond to the voltage shifts measured with the corresponding gates (colored gates are plotted along the $x$ axis at their actual spatial cordinate). The red spline connecting the measured points is a guide to the eye. Notably, the bright state charge is spread rather homogenously along the suspended nanotube. **c.** Similarly measured spatial distribution of the dark state added charge, $\rho_{ND}(x)$, observed to be concentrated at the center of the confinement well.

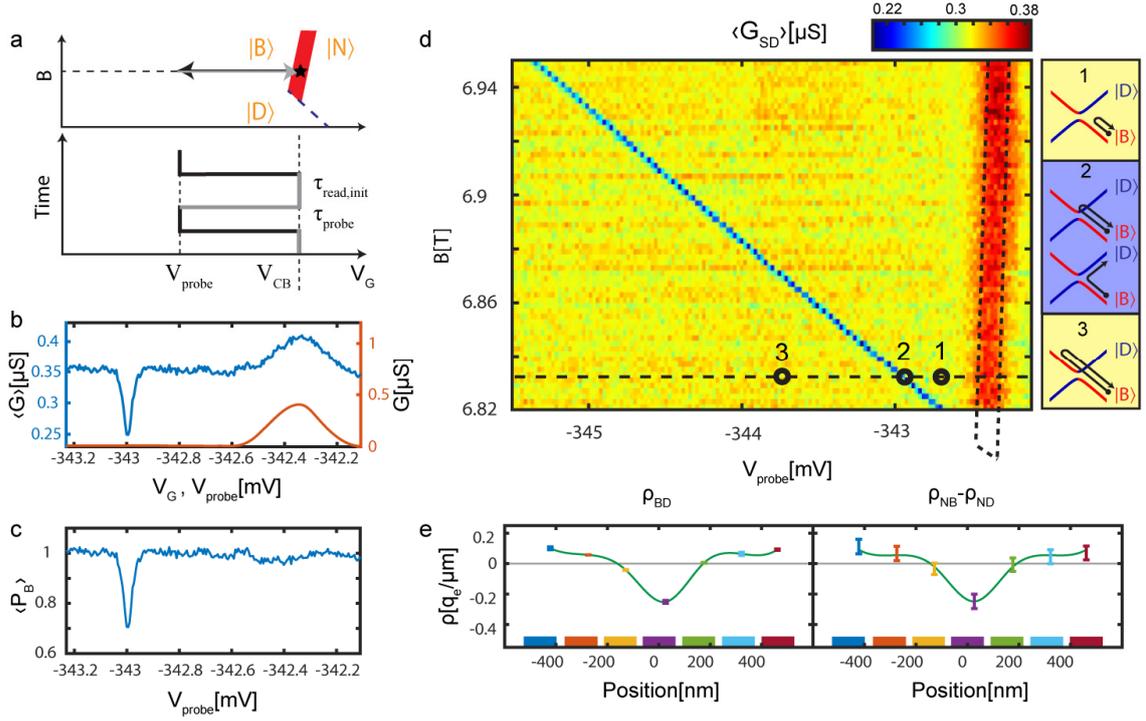

**Figure 3: Time domain measurements of the qubit transition a**. Measurement sequence, plotted in gate voltage – $B_{\parallel}$ plane (top) and in gate-voltage time diagram (bottom). The state is initialized on the $|N\rangle \leftrightarrow |B\rangle$ Coulomb blockade (CB) line (red, star marks initialization point). Voltage on the center three gates is rapidly ramped to a value $V_{probe}$, dwelling a time $\tau_{probe}$, and then ramped back to the original point on the CB line, dwelling a time $\tau_{read\_init}$. The sequence is repeated periodically. **b.** Conductance averaged over the above sequence, $\langle G \rangle$, measured as a function of $V_{probe}$ (blue) and the quasi-DC conductance, $G$, measured at the same point in $V_G$. **c.** Bright state return probability, $\langle P_b \rangle$ as a function of $V_{probe}$, determined from $\langle G \rangle$ and $G$ in panel b (see text). **d.** Similar measurement of $\langle G \rangle$ but now as a function of $V_{probe}$ and $B_{\parallel}$. A red peak is visible at the location of the Coulomb blockade peak and a much narrower blue dip appears along a line that corresponds to the $|D\rangle \leftrightarrow |B\rangle$ crossing. Both a fast traverse to a point before this line (marked 1) or after this line (marked 3) do not alter the state of the system from its initialized $|B\rangle$ state, however, a fast traverse to a point on the line (marked 2) results in occupation of the $|D\rangle$ state with a significant probability. The corresponding traverses in energy are shown in the side panels. **e.** Left: The change in spatial charge distribution at the $|D\rangle \leftrightarrow |B\rangle$ transition, $\rho_{BD}(x)$, imaged using gate resolved capacitance shift imaging of this line (as outlined in Fig. 2a). Right: The difference between the the spatial charge distributions measured at the $|N\rangle \leftrightarrow |B\rangle$ and $|N\rangle \leftrightarrow |D\rangle$ transitions, $\rho_{NB}(x) - \rho_{ND}(x)$, taken from figs. 2c and 2d, strongly resembling the directly imaged $\rho_{BD}(x)$ in the left panel, further confirming that the narrow line corresponds to the $|D\rangle \leftrightarrow |B\rangle$ transition.

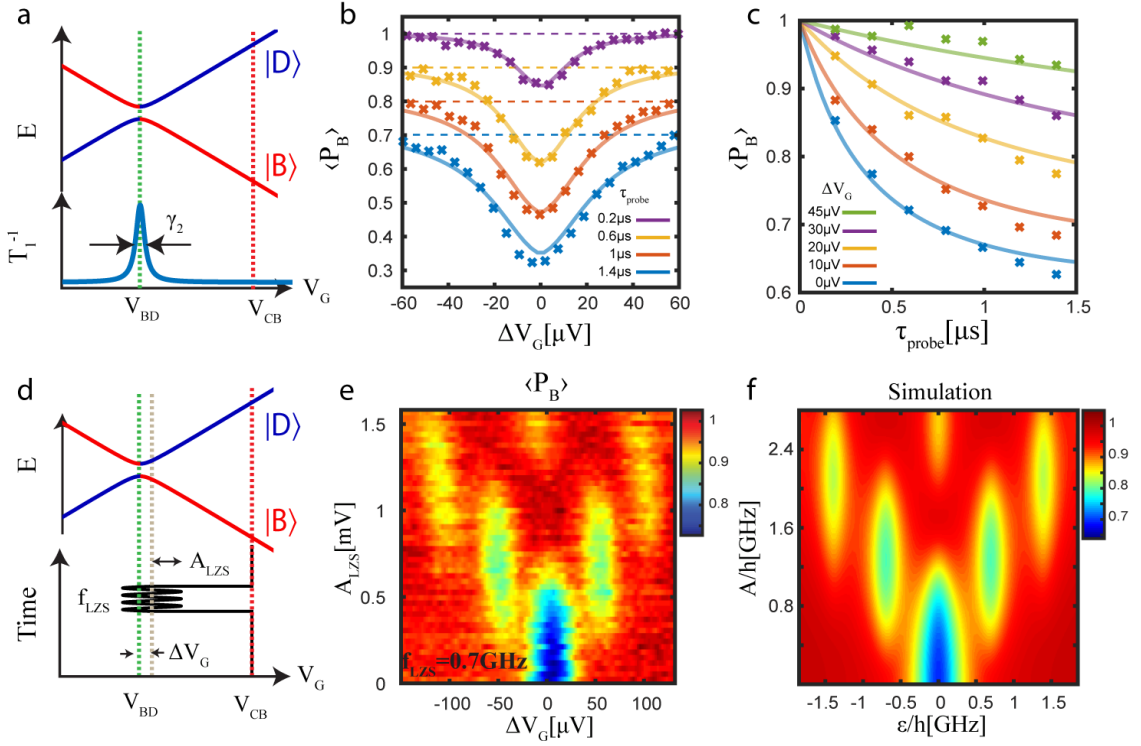

**Figure 4: Measuring the decay and coherence times of the atomic-like qubit. a.** Incoherent transition between the two states of the qubit, $|B\rangle$ and $|D\rangle$, occurs near their degeneracy point in gate voltage, $V_{BD}$. **b.** Measured $\langle P_B \rangle$ as a function of $\Delta V_G = V_G - V_{BD}$, for different $\tau_{probe}$. Curves are offseted for clarity, and the dashed colored lines corresponds to $\langle P_B \rangle = 1$ for the corresponding traces. **c.** Measured $\langle P_B \rangle$ as a function of $\tau_{probe}$ for few values of $\Delta V_G$. Solid lines in both panels b and c are fits to a model in which the decoherence results from charge noise, described in the main text. All the lines are calculated with a single set of parameters. **d.** Landau-Zener-Stuckelberg (LZS) interferometry of the qubit in which a modulation $A_{LZS}\sin(2\pi f_{LZS}t)$ is added to the time sequence in panel a during the probing stage. **e.** $\langle P_B \rangle$ measured as a function of $\Delta V_G$ and $A_{LZS}$, with $f_{LZS} = 0.7 GHz$, exhibiting the LZS interference pattern. **f.** Simulation reproducing the measurements in panel e (see text), using the same parameter used to reproduce the data in panels b and c.

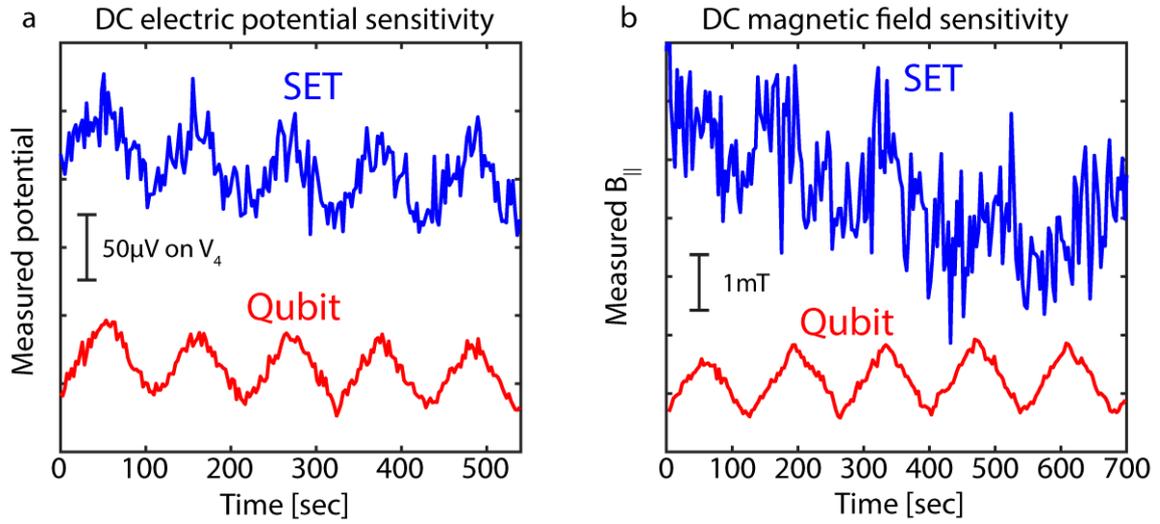

**Figure 5: Comparing SET and qubit performance in sensing DC electric potential and DC magnetic fields. a.** DC electric potential meansurement: we slowly ramp the potential of a single gate ($V_4$) up and down (sawtooth) with a period of ~100 seconds, and probe it using the device in SET modality (blue) or by detecting qubit state transitions using the time domain sequence described in Fig.3a (red). The potential scale is shown by the vertical bar. **b.** Similarly, but now for DC magnetic field measurement. In both quantities, we observe a significant improvement in sensitivity when the device operates in qubit modality rather than SET modality.

# Supplementary Information

## Electric and Magnetic Field Nano-Sensing Using a New, Atomic-like Qubit in a Carbon Nanotube

I. Khivrich, S. Ilani

## Contents

# S1. Full set-up description

The experiment is performed in a cryogen-free dilution refrigerator. The gate voltages $V_1..V_7$ are controlled by independent DC sources. In addition, fast pulses are introduced on gates 3,4,5 using a Tabor 2184 AWG through bias T connections at the cryogenic stage.

The experiment requires highly sensitive measurement of conductance through the device. However, commonly used acoustic frequency lock-in measurements are contaminated due to mechanical vibration of the fridge through triboelectric effect in the cables[1]. In order to avoid this issue, we measure the conductance through the nanotube G at a higher (~1.45MHz) frequency using the scheme detailed in Fig. S1.

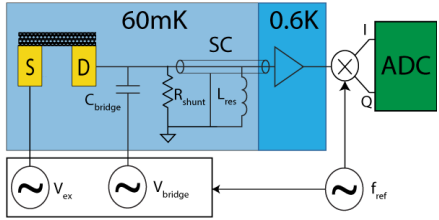

**Figure S1. Conductance measurement method** The measurement is performed at $f \sim 1.45 MHz$ defined by the resonance frequency of an LC circuit formed by an inductor ($L_{res}$) and the capacitance of a superconducting cable (SC) through a custom-built two-stage HEMT amplifier. The excitation signal $V_{ex}$ is applied on the source electrode of the device, while a separately controlled compensating signal $V_{bridge}$ is applied on the drain in order to null direct capacitive crosstalk. The measurement result is demodulated to near baseband and sampled (ADC).

# S2. Gate-based charge density imaging

## 1. General approach

The gate-based imaging experiments described in Fig 2c, 2d and 3e in the main text give a spatial picture of how the local density along the nanotube changes as we cross transitions between the three relevant states in our experiment: $|N\rangle, |B\rangle$ and $|D\rangle$. These states have well defined spatial charge density distributions: $\rho_N(x), \rho_B(x), \rho_D(x)$, and across the different transition lines we measure the difference of the corresponding densities. For example across the $|N\rangle \leftrightarrow |B\rangle$ transition our experiment images $\rho_B(x) - \rho_N(x)$. Below we explain why this is the result obtained from the capacitance measurements to individual gates as is shown in the main text. The spatial potential induced along the nanotube by biasing gate $i$ by $\Delta V$, $\phi_i(x)$, can be calculated using electrostatic simulations, which as we have shown in past experiments[2,3] describe quantitatively well the measured potentials. This extra potential will lead to an energy shift of the coulomb peak (in the case of the $|N\rangle \leftrightarrow |B\rangle$ and $|N\rangle \leftrightarrow |D\rangle$ transitions) or to and energy shift of the qubit transition (on the $|D\rangle \leftrightarrow |B\rangle$ line) which is directly proportional to the local density above the gate $\delta E_i^\beta = \int dx\, \rho_\beta(x)\phi_i(x)$, $\beta = |N\rangle, |B\rangle, |D\rangle$). The crossing between states $\beta_1, \beta_2$, will thus shift in energy by $\delta E_i^{\beta_1,\beta_2} = \delta E_i^{\beta_1} - \delta E_i^{\beta_2}$. In the experiment we measure the the global voltage shift $\delta V_i^{\beta_1,\beta_2}$ that nulls the energy perturbation introduced by changing the voltage on a single gate $i$ by $\Delta V$, where the global voltage is applied on a subset of the gates defined by the vector $v = (1,1,1,1,1,1,1)$ for the measurements across the Coulomb transitions and $v = (0,0,1,1,1,0,0)$ for measurements across the qubit transition, which are done only with the three central gates. We thus obtain,

$$\delta V_i^{\beta_1,\beta_2} \sum_j v_j \left(\delta E_j^{\beta_1} - \delta E_j^{\beta_2}\right) = \Delta V \left(\delta E_i^{\beta_1} - \delta E_i^{\beta_2}\right) \tag{S1}$$

I.e.,

$$\rho_{\beta_1}(x_i) - \rho_{\beta_2}(x_i) = \frac{\delta V_i^{\beta_1,\beta_2}}{\Delta V} N \tag{S2}$$

Where N is a normalization constant independent of $i$.

## 2. Imaging charging lines

The shifts along the global voltage axis of the bright charging line ($\delta V_i^{NB}$) are found as the shift of the center of the Coulomb blockade peak ($|N\rangle \leftrightarrow |B\rangle$) for each perturbed gate, i.e. global voltage shifts required to return the charging condition (Fig. S2a,b). In order to find the global voltage shifts for the dark transition line, we derive its location by finding both the shift of the Coulomb blockade peak ($\delta V_i^{NB}$), and the shift of the boundary between N, D states along the magnetic field axis ($\Delta B_i$). The latter is found by following the decrease of conductance at constant $V_G$, and estimating the crossing point using a linear fit. The resulting point in ($V_G, B_{||}$) is taken as the upper corner of the dark charging line (white dashed line in Fig. S2,a). The shifts of this line in the vicinity of the upper corner are insensitive to the shifts of the lower corner, and can be estimated as $\delta V_i^{ND} = \delta V_i^{NB} + \Delta B_i \alpha$, where $\alpha$ is the measured unperturbed slope of the dark charging line (slope of blue line in Fig.1f). We checked that this method of following the dark charging line (dashed white line in Fig. S2) gives identical results to finding the position of this line by following the line connecting the two bright lines endpoints that this dark line should connect.

For a charging line (i.e., steady state conductance measurement), we use as the global gate voltage axis a vector of the form $v = (1,1,1,1,1,1,1)$. Since for a charging line, $\int (\rho_B(x) - \rho_N(x)) = 1|q_e|$, and for the values of $\phi_i$ obtained in using finite element simulation, $\sum \phi_i(x)$ is flat over most of the length of the nanotube (blue curve in Fig. S3a), and $N$ is roughly independent on $\rho$. As a result, we use the (S2) with proper normalization to produce an estimate of $\rho_A(x_i) - \rho_B(x_i) \sim \frac{\delta V_i^{AB} q_e}{\Delta V d}$ where $d$ is the distance between the gates. This estimate ignores the significant crosstalk between the neighboring gates, and a more correct estimation, assuming a concrete model for charge distribution, is explained in section S3.

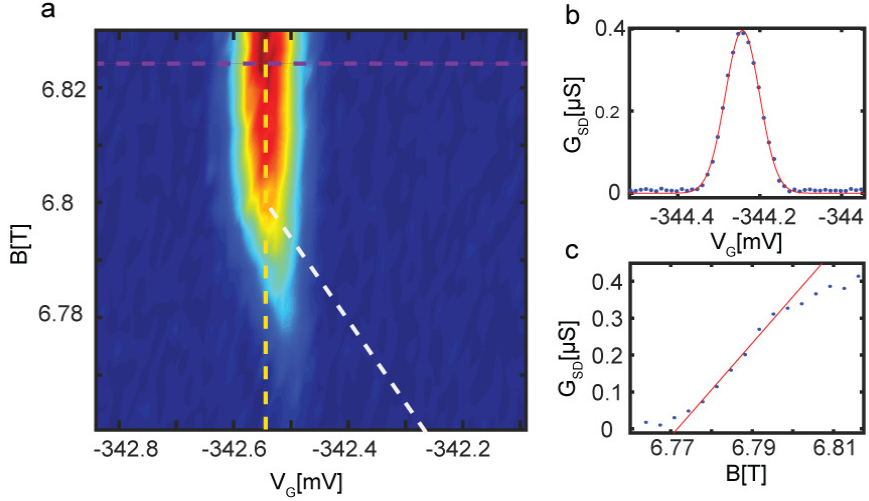

**Figure S2. Imaging charge density changes on transition lines** The location of the Coulomb blockade on $V_G$ (a, yellow line) is found by Gaussian fitting the transport at a higher B (b). The triple point where N, B, D are degenerate is located as middle point of conductance along the yellow line from a linear fit to the slope (c). The calculated N-D boundary (white line) is drawn from the triple point at a constant slope.

### 3. Imaging BD transition line

Two key differences compared to the previous case are a different assumed normalization ($\int dx(\rho_{NB} - \rho_{ND}) = 0$), and a different compensation vector: $v = (0,0,1,1,1,0,0)$ due to AWG outputs not being connected to all the gates.

The shift in the location of BD transition line after perturbation $\delta V_i^{BD}$ is measured as a sum of the two contributions: the shift in the Coulomb blockade peak used for readout, and the shift of the BD line with respect to the readout point. The normalization parameter, $N = \sum v_j(\delta E_j^B - \delta E_j^D)$, is estimated by imaging the two charging lines (S2.2)

## S3. Determining the qubit charge redistribution

The spatial extent of gate potentials $\phi_i(x)$ creates significant overlaps of sensitivity between different gates, resulting in smearing of the qubit charge redistribution $\rho_{BD}(x)$ measured via the capacitance to the gates (Fig 3e, main text) as compared to the actual $\rho_{BD}(x)$. We can determine the spatial distribution of the potential produced by each one of the gates, $\phi_i(x)$, using finite element electrostatic simulations (Comsol) (Fig. S3a). These calculated $\phi_i(x)$ were checked in our previous experiments against the actual potentials measured by scanning single electron transistor imaging, and where found to be accurate to within 10%. From Fig S3a, it is clear that a gate is capacitively coupled not only to the nanotube segment above it, but has also substantial coupling to positions further away along the nanotube. To deconvolve the effects of these overlaps and obtain a better estimate of $\rho_{BD}(x)$, we assume a simple form for $\rho_{BD}(x)$ that captures the essential shape of the charge redistribution: $\rho_{BD}(x) = A_1 \exp\left(-\left(\frac{x}{w_1}\right)^2\right) + A_2 \exp\left(-\left(\frac{x}{w_2}\right)^2\right)$. By optimizing the parameters $A_1, A_2, w_1, w_2$, we reproduce the measured gate induced shifts, $\delta V_i = \int \rho_{BD}(x)\phi_i(x)dx$. A correspondence is observed between the reproduced $\delta V_i$ for the optimal values of the parameters (+, Fig. S3b) and the measured values, shown using confidence intervals in Fig. 3b. As expected, the reconstructed charge density distribution (blue, Fig. S3c) is narrower than the curve drawn by the measured $\delta V_i$ (green, Fig. S3c), since the latter includes the smearing by the cross capacitances of each gate to adjacent locations along the nanotube. The reconstructed shape, which represent more accurately the qubit charge density redistribution, has a $FWHM \sim 100 nm$. The overall charge change in the central node (150$nm$ width, bounded by dashed lines) is $\delta q \sim 0.1 q_e$.

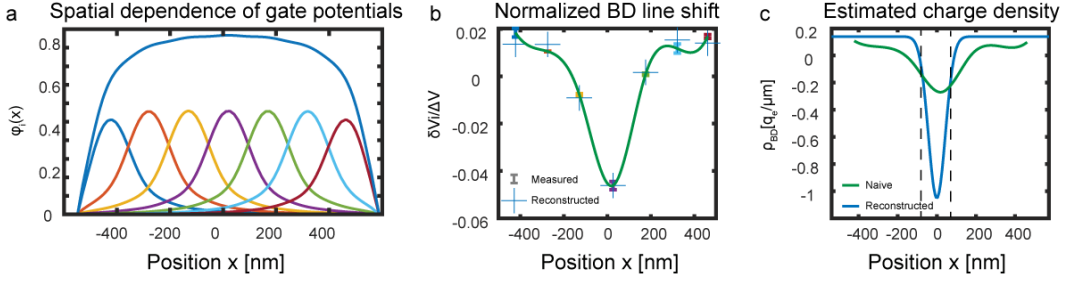

**Figure S3. Reconstructing charge density change on BD transition** (a) Calculated potentials induced by the gates, normalized to voltage applied on the gates. The overlap between the adjacent gates is significant, and the naïve estimation of charge distribution using a single gate is quantitatively imprecise. We model the density change as a sum of two Gaussians, and fit their parameters so that the modelled values obtained by smearing using the gate potentials reproduces the measured values (+ signs and error bar intervals in b). The resulting reconstructed distribution (blue, c) is indeed significantly compressed as compared to naïve calculation of the density (green, c), allowing a more precise estimation of sensitivity for optimally located sources.

## S4. Theoretical description of decoherence rate

We describe the behavior of the system in Bloch equation formalism, in the manifold of the two nearly degenerate states in the far-detuned basis ($|B\rangle, |D\rangle$). The unitary evolution is described by the Hamiltonian[4]

$$H = \frac{\epsilon(t)}{2} \sigma_z + \frac{\Delta}{2} \sigma_x \quad (S3)$$

Where the detuning $\epsilon(t)$ is steered in time by fast changes of the gate voltage $V_G$, and the tunneling rate $\Delta$ is constant. The dominant decoherence mechanism results from noise in $\epsilon$ leading to decay of the Bloch vector to the Z axis with rate $\gamma_2$. In addition, a less significant noise in $\Delta$ leads to decay to the XY plane with rate $\gamma_1$. The steady state thermal occupation vector for each detuning $\epsilon$ is given by

$$S_{th}(\epsilon) = \left(0, 0, \tanh\left(\frac{\sqrt{\epsilon^2 + \Delta^2}}{2T}\right)\right)^T \quad (S4)$$

Where T is the temperature of the environment. The full description of the evolution of the Bloch vector is

$$\dot{S} = M(\epsilon) S + \gamma_1 S_{th}(\epsilon) \quad (S5)$$

Where

$$M(\epsilon) = \begin{pmatrix} -\gamma_2 & \epsilon & 0 \\ -\epsilon & -\gamma_2 & \Delta \\ 0 & -\Delta & -\gamma_1 \end{pmatrix} \quad (S6)$$

The effective decay rate $T_1$ is estimated by solving the equation at steady state ($\dot{S}_x = \dot{S}_y = 0$), and obtaining $\dot{S}_z = -\frac{S_z}{T_1(\epsilon)} + const$:

$$T_1^{-1}(\epsilon) = \gamma_1 + \frac{\Delta^2}{\gamma_2 + \frac{\epsilon^2}{\gamma_2}} \tag{S7}$$

Thus the expected width of the observed transition line is determined by the dephasing rate $\gamma_2$, rather than temperature, as illustrated in Fig. 4 of the manuscript.

## S5. Full calculation: self-consistent occupation and observed current

In a naive interpretation, the dependence of the observed mean occupation $\langle P_B \rangle$ on $\tau_{probe}$ should be an exponential decay to a thermal equilibrium value. However, this interpretation neglects the possibility of incomplete initialization of the system to its ground state $|B\rangle$ at the end of the read step, which will also depend on the detuning $\Delta V_G$ and timing parameters $\tau_{read,init}, \tau_{probe}$. Thus, in order to quantitatively interpret the dynamical behavior of the system, we need to solve a more detailed model.

Our measurement consists of repeating two steps with different values of detuning $\epsilon_{probe}$, $\epsilon_{read}$ for the durations $\tau_{probe}, \tau_{read,init}$, hence the evolution of $S$ over a single period of the measurement is given by

$$S' = e^{M(\epsilon_{probe})\tau_{probe}} \left( e^{M(\epsilon_{read})\tau_{read,init}} (S - S_{th}(\epsilon_{read})) + S_{th}(\epsilon_{read}) - S_{th}(\epsilon_{probe}) \right) + S_{th}(\epsilon_{probe}) \tag{S8}$$

At steady state, we solve for $S' = S$, obtaining the value of the Bloch vector at the beginning of the read step:

$$S_0 = \left( e^{M(\epsilon_{probe})\tau_{probe}} e^{M(\epsilon_{read})\tau_{read,init}} - I \right)^{-1} \tag{S9}$$

$$= \left( e^{M(\epsilon_{probe})\tau_{probe}} \left( S_{th}(\epsilon_{read})(I - e^{M(\epsilon_{read})\tau_{read,init}}) - S_{th}(\epsilon_{probe}) \right) + S_{th}(\epsilon_{probe}) \right)$$

Integrating over the read step (see Fig. S4), we get the reduction of the measured conductivity $\langle G \rangle$ with respect to transport value $G_0$:

$$m := \frac{\langle G \rangle}{G_0} = 1 - \frac{(1 - S_{0,z})}{2\gamma_1 \tau_{read}} (1 - \exp(-\gamma_1 \tau_{read})) \tag{S10}$$

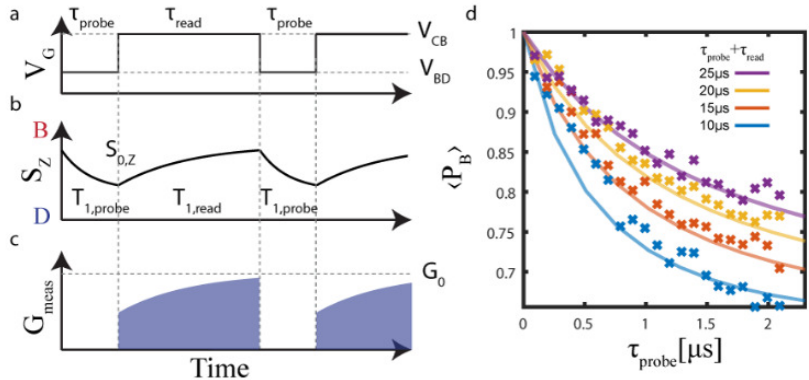

**Figure S4: accurate interpretation of the observed conductance** During the read period (a), the occupation (b) is initialized towards ground state B (low $S_z$), while during the probe period, the probability of excited state D can increase ($S_z$ decreases). The current through the system is only observed during the read period (c), and its value represents the average occupation of B during this period (shaded curves). The resulting rate equations explain the observed conductance at $\Delta V_G = 0$ for a range of $\tau_{read,init}, \tau_{probe}$ (d).

Substituting (4), (6), (9) into (10), we obtain an expression for the observed change in the conductance $\langle G \rangle / G_0$ as a function of three free parameters $\gamma_1, \gamma_2, \Delta$, while the electron temperature $T$ and the conversion ratio of $\Delta V_G$ to energy scale ($\epsilon$) are found in independent measurements. By simultaneously fitting the model to the experimental data displayed in Fig. 4b,c and Fig. S4d (continuous lines) we obtain the values $\gamma_1 = 2\pi \times (1.5 \pm 0.15) kHz$, $\gamma_2 = 2\pi \times (185 \pm 6) MHz$, $\Delta = 2\pi \times (2 \pm 0.12) MHz$, where the confidence intervals are $2\sigma$ and are obtained using Monte Carlo approach.

To obtain a more useful formula, an approximate solution of the same problem can be found by looking at the evolution of z component of the Bloch vector. Two effective values of the decay rate (7) are used: $T_{1,read}^{-1} = T_1^{-1}(\epsilon_{read})$, $T_{1,probe}^{-1} = T_1^{-1}(\epsilon_{probe})$. In this case, the self-consistent relation (9) can be written as

$$S_{0,z} = \frac{S_{th,z}\left(e^{\frac{\tau_{probe}}{T_{1,probe}}} - 1\right)}{e^{\frac{\tau_{probe}}{T_{1,probe}}} - e^{-\frac{\tau_{read,init}}{T_{1,read}}}} \quad (S11)$$

## S6. Calibrating lever arm constants

The measurement is performed in terms of the detuning $\epsilon$ between the states $B, D$, and in order to obtain the sensitivity of electric and magnetic field measurements, we need to calibrate the electrostatic lever arm factors $\alpha_B, \alpha_D = q_e^{-1} \partial E_{B,D} / \partial V_G$ where $\delta V$ is a voltage of interest (in the following discussion, common voltage on three central gates 3,4,5), and the magnetic moments $\mu_B, \mu_D = \partial E_{B,D} / \partial B_{||}$. Using the measured slopes of red and blue lines in Fig. 1f ($S_B = \mu_B / \alpha_B$, $S_D = \mu_D / \alpha_D$), and the slope of the interface line in Fig. 3d $K = (\mu_B - \mu_D)/(\alpha_B - \alpha_D)$, we obtain

$$\frac{\alpha_D}{\alpha_B} = \frac{1 - KS_B}{1 - KS_D} \quad (S12)$$

Using the value of $\alpha_B \sim 0.15$ measured directly from Coulomb diamond diagram (Fig.S5), we obtain $\alpha_D \sim 0.2$, $\mu_B \sim -0.08 meV/T, \mu_D \sim 1.1 meV/T$ for the states discussed in the manuscript.

The obtained lever arm factor for the qubit transition $\alpha_D - \alpha_B \sim 0.05$ is consistent with the lever arm factor independently obtained from the spacing between LZS interference fringes:

$$\alpha = \frac{2\pi \hbar f_{LZS}}{55 \mu V q_e} \sim 0.0525 \quad (S13)$$

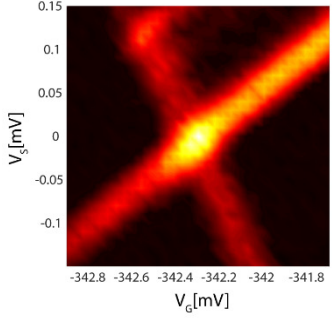

**Figure S5: Calibrating lever arm constants** The lever arm factor $\alpha_B = \partial E_B/\partial V_G$ is extracted from the Coulomb diamond measurement. Combined with the measurements in Fig. 1f and Fig. 3d, full set of values of lever arm factors $\alpha_B, \alpha_D, \mu_B, \mu_D$ is found.

## S7. Determining the qubit's potential sensitivity

We determine the qubit's sensitivity to electrical potential using the data displayed in Fig.S6. In order to estimate the measurement noise $\sigma_G$ obtained in practice, data from 6 consequtive runs in a flat region of the measurement is taken (black box in Fig. S6). By plotting the envelope calculated as mean of the measurement $\pm 1.5\sigma_G$, we verify that the samples are uniformly confined by this envelope, and the obtained noise value can be used for all values of $\Delta V_G$. Using $\sigma_G$, the slope (dashed line) and the lever arm factor (S6), we find the sensitivity to detuning to be $\delta\epsilon = 60 neV/\sqrt{Hz}$.

The shift of detuning due to a local electrical potential $\phi(x)$ is given by $\delta\epsilon = \int \phi(x)\rho_{BD}(x)$. As is apparent from figure S3c, the qubit charge redistributes from the center to the sides ($\rho_{BD}(x)$ is negative in a region of width $W \sim 100nm$ in the center and positive on the sides). This means that the qubit will be insensitive to potentials whose wavelengths are much larger than $W$, making it immune against far field potential noise. The optimal sensitivity in potential measurements will be obtained when the measured feature is comparable in size to the resolution that is set by $W$. In this case the potential sensitivity can be determined to be $\delta V = \frac{\delta\epsilon}{\delta q} \sim 600 nV/\sqrt{Hz}$, where $\delta q$ is the redistributed charge (S3). Note that the sensitivity to voltage on gate 4, $\delta V_4 \sim 1.4 \mu V$ is about two times worse as the potential that this gate produces, $\phi_4(x)$, is smeared on a scale larger than $W$.

## S8. Theoretical limits on the potential sensitivity

The potential sensitivity obtained in the experiment (section S7), although extremely good, can still be improved if few limiting factors in the current experiment are improved. To get a sense of what is the optimal performance expected for this qubit based on its measured properties, we analyze its theoretical sensitivity limits below:

The measured quantity in our experiment is the dark vs. bright state occupation near the qubit transition. This quantity depends very sharply on the detuning, $\epsilon$, between these states. The detuning can be modified by a modulation of the magnetic field ($\delta\epsilon = (\mu_B - \mu_D)\delta B_{||}$) where $\mu_B, \mu_D$ are the magnetic moments of the two states, or by electric potential ($\delta\epsilon = (\alpha_B - \alpha_D)\delta V$), where $\alpha_B, \alpha_D$ are the lever arm factors of the gate capacitance to the charge distributions of the states. The detuning is detected through the modulation of current $\delta I = I_0 m(\delta\epsilon)$ where $I_0$ is the readout current measured on Coulomb blockade peak. The sensitivity of $m$ to $\epsilon$ can be seen from S7, S11 to be of the form $\frac{\partial m}{\partial \epsilon} \sim A\hbar\gamma_2$, where $A$ is a numerical factor depending on the timing parameters and approaching the value of $\sim 10$ when optimized. The signal to noise ratio for measuring $m$, $SNR_m$ in our case is limited by our contact resistance, $R_{contact}$. In such a case of resistive contacts, $SNR_m$ is determined by the amplifier noise floor $i_n \sim 25 fA/\sqrt{Hz}$ ($SNR_m = \frac{I_o}{i_n}$, where $I_0$ is limited by the maximal AC excitation for conductance measurement $V_{ex} \sim 2.3 k_B T/q_e$ and the device contact resistance $R_{contact}$). In total, for this case, the sensitivity to detuning between the energies of B,D is:

$$\delta\epsilon_{contact\ resistance\ limited} \sim \frac{4\hbar\gamma_2 i_n q_e R_{contact}}{k_B T} \quad (S14)$$

For vertex properties described in the manuscript, $\delta\epsilon \sim 45 neV/\sqrt{Hz}$, which is translated to $\delta V \sim 450 nV/\sqrt{Hz}$.

The device in this study had a far from optimal contact resistance, $R_{contact} \sim 2M\Omega$. In principle this resistance could be improved down to tens of $k\Omega$ range, improving the sensitivity of the described method. Ultimately, the signal to noise of measuring $m$ will be limited by the shot noise of the measurement and will become insensitive to $R_{contact}$ (since the current during the readout stage assumes one of two possible values): $SNR_m = 1/\sqrt{\tau_{read,init} + \tau_{probe}}$. For a finite bandwidth $BW$ for setting the gate voltages determining the minimal $\tau_{probe} \sim BW^{-1}$, the optimal sensitivity is obtained for $\tau_{read} \sim \tau_{probe} T_1^{probe}/T_1^{read}$, hence the optimal value is $SNR_m = \frac{\sqrt{\gamma_1 \gamma_2 BW}}{\Delta}$. Thus, the ultimate bound for sensitivity is

$$\delta\epsilon_{shot\ noise\ limited} \sim \frac{10\sqrt{\gamma_2}\hbar\Delta}{\sqrt{\gamma_1}\sqrt{BW}} \quad (S15)$$

For the device equivalent to those presented in the manuscript and $BW \sim 1 GHz$ $\delta\epsilon_{shot\ noise\ limited} \sim 4 neV/\sqrt{Hz}$, which translates to $\delta V \sim 40 nV/\sqrt{Hz}$ and $\delta B \sim 3.5 \mu T/\sqrt{Hz}$.

The factors leading to the difference between the naive estimation of the expected performance (S6) and the obtained result include slow drifts of device conductance, which can be avoided performing duty cycle measurements. In addition, since the device is sensitive to both magnetic and electric field, slow drifts in local electric field will contribute to measurement noise of magnetic field and vice versa. The device can be optimized for a required type of measurement by choosing other working points in $(V_G, B)$ space, which may have different coupling differences $(\alpha_B - \alpha_D)$, $(\mu_B - \mu_D)$. For instance, in Fig. S6b we choose a vertex with significantly smaller $\alpha_B - \alpha_D$, improving the resulting magnetic field sensitivity to $39 \mu T/\sqrt{Hz}$ (evaluated in the same way). Our sensing relies on the coherence limited transition of the qubit, and in this sense is similar to relaxometry measurements in NV centers, used to probe high frequency noise. So far we could not use the full potential of coherent manipulation of the qubit, due to the rather short $T_2^*$. With improvement in $T_2^*$ it should be possible to use dynamic decoupling protocols, and thus be limited by $T_2$ that is likely to be significantly longer. In this case further improvement to the sub $nV$ potential sensitivity might become possible, as was demonstrated in singlet-triplet qubits in GaAs double quantum dots[5].

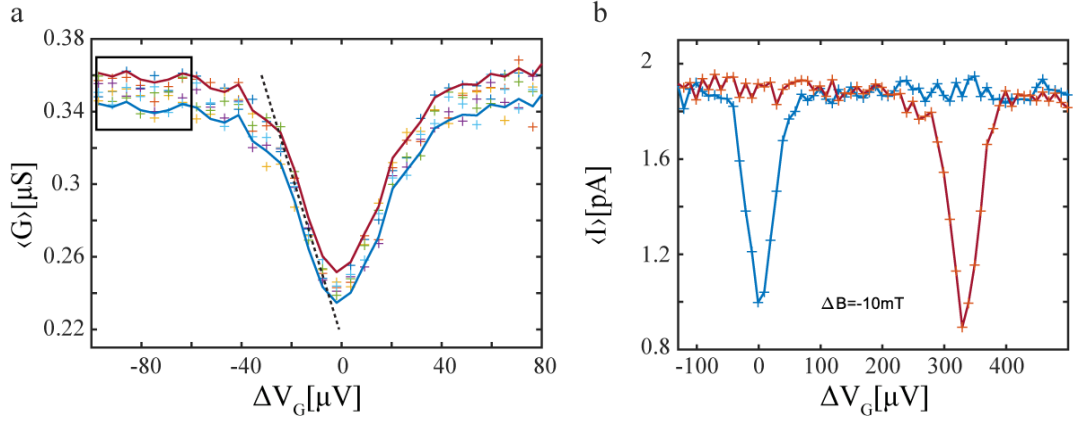

**Figure S6: estimating detection sensitivity** a. Noise statistics are estimated from 6 consecutive scans (+); noise power $\sigma_G$ extracted in a far-detuned region (black box) is seen to be uniformly correct for all detuning values from the envelope curves, separated by $3\sigma_G$. The sensitivity to $\Delta V_G$ is obtained from $\sigma_G$ and the maximal slope of the conduction dip (dashed line) b. Choosing a vertex with smaller difference between the charge distributions of $|B\rangle, |D\rangle$ leads to broadening of the conduction dip and increase in $B_{||}$-dependence of its location, shown in two graphs for slightly different magnetic field values. Overall, the sensitivity to $B_{||}$ is improved.

## S9. Bright-dark transition lines for a range of $B_{||}$

In order to implement the demonstrated measurement scheme, the control parameters of the system, $V_G$ and $B_{||}$ need to be tuned to obtain a dark state with sufficiently long lifetime, sufficiently close to a Coulomb blockade peak to allow fast gating transitions between readout (B-N) and measurement (B-D) lines. We demonstrate that these requirements can be fulfilled over a range of values of $B_{||}$ by showing a number of observed B-D transition lines (dashed green lines in Fig. S7), shown with respect to conductance measurement of the device (electrons side). The maximal distances of the displayed lines from the Coulomb peaks represent the maximal reachable fast gate voltage transition amplitude in our set-up. The observed detuning dependence of the transition lines (blue) is shown for each case. Overall, while only some of the vertices in the transport diagram were measured, the displayed lines allow replicating the experiment over a significant fraction of $B_{||} = 3T..6T$ range. In additional measurements we have also observed the effect in other parts of the $(V_G, B_{||})$ diagram, for both electron and hole doped dots, and have seen that similar qubit transition exists at least up to $B_{||} = 8T$ (not shown).

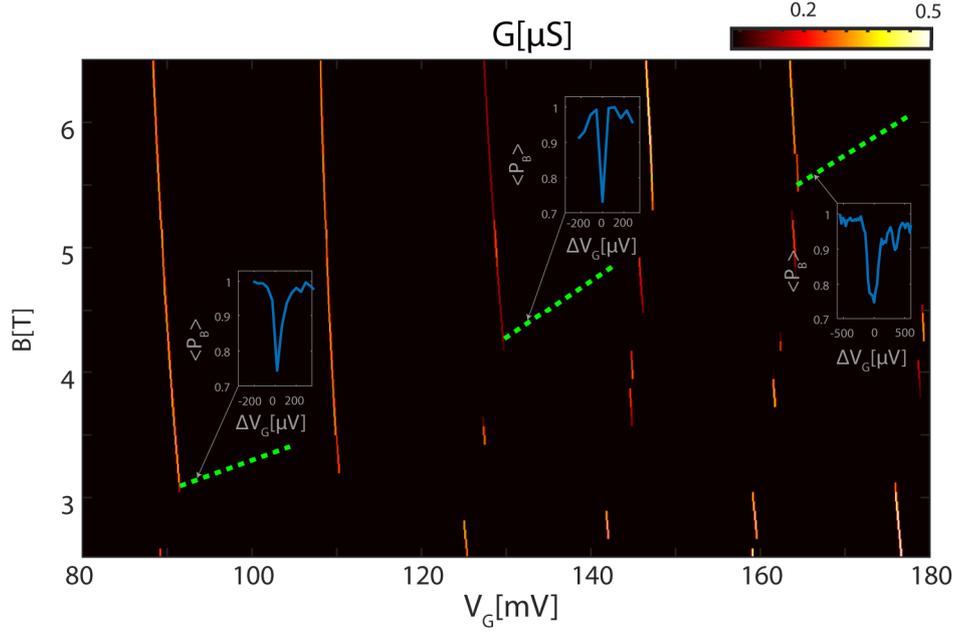

**Figure S7: bright-dark transition lines for a range of $B_{||}$** Zig-zag pattern of Coulomb blockade(orange colormap). Measured locations and slopes of lines of transition to dark state obtained using the time-domain method shown in green dashed lines, terminating at distance from CB peak limited by maximal fast gating jump amplitude (setup - limited). The insets show the observed $\langle P_B \rangle (\Delta V_G)$ (equivalent to Fig. 3c). In some cases, these lineshapes are more complex than the simple case described in the manuscript, and indicate the existence of other states in addition to N,B,D. Despite this complexity, the observed magnetic moments and line sharpness are similar to the simple qubit described in the paper, thus allowing similar measurement sensitivity. Overall, using the displayed lines, the electric and magnetic sensing mechanism can be used for most of the values of $B_{||}$ between $3T$ and $6T$ (We have measured similar lines up to $8T$, not shown).

## S10. LZS simulation

The simulation is done using QuTip package[6]. We simulate a system described by Hamiltonian (S3) where $\epsilon(t)$ is modulated harmonically, and the decoherence operators $\sqrt{\frac{\gamma_1}{2}}\sigma_-$ and $\sqrt{\frac{\gamma_2}{2}}\sigma_z$. While a qualitatively similar result can be obtained by finding the steady state $\langle P_B \rangle$ of the resulting Lindblad equation, a better quantitative match is obtained by calculating the steady state occupation for two consecutive propagation steps, in which $\epsilon(t) = \epsilon_0 + A_{LZS}\sin(\omega t)$ for time $\tau_{probe}$, and $\epsilon(t) = \epsilon_{CB}$ for time $\tau_{read,init}$.